\documentclass[11pt, a4paper, logo, copyright, nonumbering]{lti}
\usepackage[authoryear, sort&compress, round]{natbib}
\usepackage{dblfloatfix}
\usepackage{ulem}
\usepackage{caption}
\usepackage{dramatist}
\usepackage{xspace}
\usepackage{pifont} 
\usepackage{multirow}
\usepackage{tcolorbox}
\tcbuselibrary{skins}
\usepackage{xltabular}
\usepackage{longtable}
\usepackage{amsfonts}
\usepackage{amsmath}
\usepackage{amssymb}
\usepackage{lineno}
\usepackage{multirow}
\usepackage{adjustbox}

\usepackage[bottom]{footmisc}

\usepackage{CJKutf8}
\usepackage{subfigure}
\usepackage{setspace}

\usepackage{dsfont}
\usepackage{array}
\usepackage{tabularx}
\usepackage{subfigure}
\usepackage{xcolor}
\usepackage{wrapfig}
\usepackage{booktabs}

\definecolor{ltired}{HTML}{780000}

\usepackage{lipsum}
\usepackage{multicol}
\usepackage{makecell}

\usepackage{microtype}
\usepackage{graphicx}
\usepackage{csquotes}
\usepackage{xurl} 
\usepackage{listings}
\lstdefinestyle{prompt}{
  basicstyle=\ttfamily\tiny,
  breaklines=true,
  breakatwhitespace=true,
  postbreak={}
}
\lstdefinestyle{python}{
  language=Python,
  basicstyle=\ttfamily\tiny,
  breaklines=true,
  breakatwhitespace=false,
  postbreak={},
}

\newtcolorbox[auto counter, number within=section]{positionbox}{
  enhanced,
  colback=gray!10,
  colframe=black,
  boxrule=0.6pt,
  arc=3pt,
  left=8pt,
  right=8pt,
  top=12pt,
  bottom=8pt,
  coltitle=black,
  title={\textbf{Position}},
  attach boxed title to top left={yshift=-8pt, xshift=10pt},
  boxed title style={
    enhanced,
    colback=gray!30,
    colframe=black,
    coltitle=black,
    boxrule=0.6pt,
    arc=3pt,
    left=4pt,
    right=4pt,
    top=2pt,
    bottom=2pt
  }
}

\usepackage{amsthm}
\theoremstyle{plain}

\theoremstyle{remark}

\usepackage[disable]{todonotes}

\newcommand{\df}[1]{}
\newcommand{\gn}[1]{}
\newcommand{\diyi}[1]{}
\newcommand{\zw}[1]{}
\newcommand{\kl}[1]{}
\newcommand{\kn}[1]{}
\newcommand{\john}[1]{}
\newcommand{\vc}[1]{}
\newcommand{\zijian}[1]{}
\newcommand{\dfcomment}[1]{}
\newcommand{\gncomment}[1]{}
\newcommand{\diyicomment}[1]{}
\newcommand{\zwcomment}[1]{}
\newcommand{\klcomment}[1]{}
\newcommand{\kncomment}[1]{}
\newcommand{\johncomment}[1]{}
\newcommand{\vccomment}[1]{}
\newcommand{\zijiancomment}[1]{}
\newcommand{\kilian}[1]{}
\newcommand{\yuxiang}[1]{}

\makeatletter
\def\@BTrule[#1]{%
  \ifx\longtable\undefined
    \let\@BTswitch\@BTnormal
  \else\ifx\hline\LT@hline
    \nobreak
    \let\@BTswitch\@BLTrule
  \else
     \let\@BTswitch\@BTnormal
  \fi\fi
  \global\@thisrulewidth=#1\relax
  \ifnum\@thisruleclass=\tw@\vskip\@aboverulesep\else
  \ifnum\@lastruleclass=\z@\vskip\@aboverulesep\else
  \ifnum\@lastruleclass=\@ne\vskip\doublerulesep\fi\fi\fi
  \@BTswitch}
\makeatother

\addto\extrasenglish{
}

 {\begin{list}{}%
         {\setlength{\leftmargin}{#1}}%
         \item[]%
 }
 {\end{list}}

\usepackage{color, colortbl}
\usepackage{hyperref}

\title{\centering Position: Humans are Missing from AI Coding Agent Research}

\author{\centering
\normalsize{\textbf{Zora Zhiruo Wang$^{1*}$ \quad John Yang$^{2*}$ \quad Kilian Lieret$^{3*}$ \\
Alexa Tartaglini$^{2}$ \quad Valerie Chen$^{1}$ \quad Yuxiang Wei$^{4}$ \\  Zijian Wang \quad
Lingming Zhang$^{4}$ \quad Karthik Narasimhan$^{3}$ \\
Ludwig Schmidt$^{2}$ \quad Graham Neubig$^{1}$ \quad Daniel Fried$^{1}$ \quad Diyi Yang$^{2}$}}
\\
\vspace{0.3em}
$^{1}$Carnegie Mellon University \quad $^{2}$Stanford University \quad $^{3}$Princeton University \\
$^{4}$University of Illinois Urbana-Champaign
\\
\vspace{0.3em}
\small{$^{*}$Equal Contribution}
\\
\vspace{0.3em}
\small{\texttt{zhiruow@andrew.cmu.edu, johnby@stanford.edu}}
}

\renewcommand{\phi}{\varphi}

\renewcommand{\epsilon}{\varepsilon}
\renewcommand{\imath}{\mathrm{i}}

\newlength{\restsubwidth}
\newlength{\restsubheight}
\newlength{\restsubmoreheight}
\newcommand{\rest}[2]{%
        \settowidth{\restsubwidth}{\ensuremath{#2}}
        \settoheight{\restsubheight}{\ensuremath{{}_{#2}}}
        \ensuremath{{#1\hskip 0.5pt}_{\vrule\kern2pt\parbox[b][%
        4pt][b]{\the\restsubwidth}{%
                        \ensuremath{{}_{#2}}}}}
        }

\begin{abstract}
Recent progress in AI coding agent research has led to rapid improvements in agents' ability to autonomously perform complex software engineering tasks, from editing large codebases to executing long-horizon development workflows.
As these systems make strides, however, the primary bottleneck to practical usefulness increasingly shifts away from pure task-solving capability, and toward challenges in how users communicate with, supervise, and trust agents.

In this position paper, we argue for a reorientation from autonomous to \textit{human-centered} coding agents: systems designed not only to complete tasks, but to collaborate effectively with people.
We identify four core interaction-level dimensions that characterize the human-agent task-solving loop: task alignment, verifiability, steerability, and adaptability.
Finally, we outline concrete research directions to advance these dimensions, including user-involved coding environments, comprehensive verification mechanisms, and principled measures of human-agent interaction quality.
\end{abstract}

\begin{document}

\maketitle

\begin{positionbox}
AI coding agent research is fixated on autonomous task completion, treating benchmark difficulty as a proxy for practical value.
However, as coding agents have graduated from academic papers to industry products, we hypothesize that the next meaningful breakthroughs lie \textit{not} in what agents can do solo, but in their interplay with human developers and users.
\label{position}
\end{positionbox}

\section{Introduction}
\label{sec:introduction}

Recent advances in large language models (LLMs) and agent frameworks have led to rapid progress in AI coding research.
Coding agents can now modify real codebases, resolve issues in complex repositories \citep{swebench}, and execute multi-step software engineering workflows \citep{mlebench} that were previously out of reach for automated systems.
Across benchmarks and live deployments, newer models continue to outperform earlier ones with greater autonomous execution capabilities.

\begin{figure}[t!]
\centering
\includegraphics[width=0.8\linewidth]{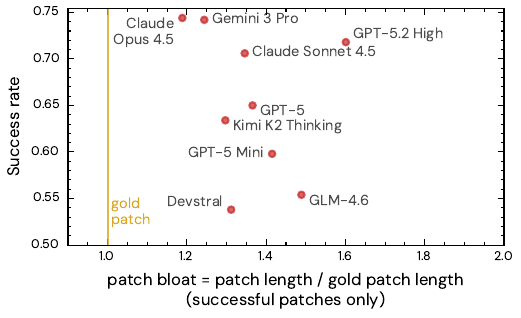}
\caption{\textbf{Agent-generated patches are consistently longer than ``ideal'' gold patches}, posing a challenge to \textit{verifiability}. In SWE-bench Verified, this pattern persists despite high success rates.}
\label{fig:teaser}
\end{figure}

However, as agent autonomy increases, the primary bottleneck to practical usefulness is shifting \citep{chen2025code}.
While earlier systems were limited by their inability to generate correct code, modern agents can produce plausible solutions, as shown by the surging scores on SWE-bench verified in \autoref{fig:teaser}. 
Empirical studies increasingly highlight failures not in producing executable code, but in misunderstanding user intent \citep{liu2023wants,jiang2022discovering}, producing outputs that are difficult to interpret or verify \citep{kumar2025developers,treude2025developers}, and behaving in ways that are hard to control or predict over time \citep{mozannar2024reading,liang2024large,chen2025code}.
For instance, agent-produced patches can be bloated and hard to verify, hindering their practical utility.
As coding agents increasingly function as general-purpose agents \citep{soni2025coding} that program to solve diverse real-world tasks beyond software engineering \citep{wang2025aiagentshumanwork}, these interaction challenges become even more pronounced.
Consequently, as agent capabilities improve, the bottleneck shifts from ``can it work?'' to ``can humans understand, trust, and work with it?'', making human-centered design essential for translating capability into practical usefulness.

This mismatch reflects a growing disconnect between agent research and real-world deployment.
Much of the field has focused on advancing autonomy, measured by success rate on harder benchmarks. Yet real-world programming is rarely a one-shot activity. Instead, it unfolds through iterative interaction, partial delegation, evolving goals, and continuous human oversight. 
In this setting, the goal of developing coding agents is not to replace human labor, but to augment complementary human strengths such as initiative and judgment. Crucially, human-centered interaction is not a temporary workaround for immature models: it enables new forms of collaboration in which agents and users jointly discover emerging workflows. As a result, utility depends not only on whether an agent can complete a task, but on whether humans can effectively work with it to shape outcomes and future work (\S\ref{sec:ai4c-today}).

\begin{figure}[t!]
\centering
    \includegraphics[width=\textwidth]{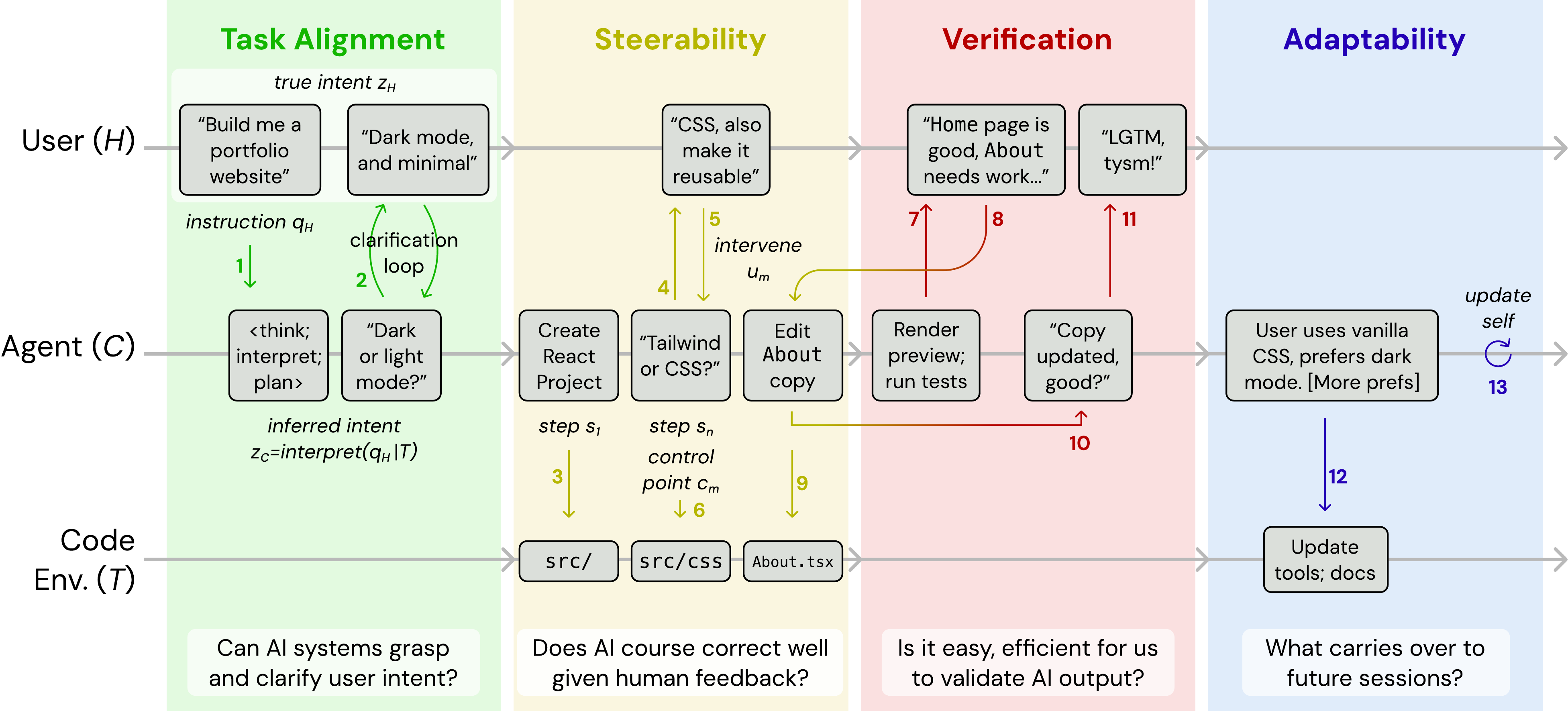}
\caption{
\textbf{The Human-Coding Agent Collaboration Loop.}
We ground our four pillars in a unified interaction loop between a user, coding agent, and code environment.
Using a concrete running example (building a personal portfolio website), we illustrate how task alignment, steerability, verification, and adaptability surface as distinct phases in a single session.
The numbers denote an order in which user/agent/codebase interactions occur.
Though depicted sequentially for clarity, these phases interleave in practice -- task alignment questions naturally may arise during steering; verification may cause an agent to revisit intent and course-correct (step 7).
}
\label{fig:what-users-want}
\end{figure}

We argue that this shift in bottlenecks calls for a corresponding adjustment in how coding agents are designed and evaluated: from maximizing agent-solo autonomy to maximizing human-centered usefulness (Position \autoref{position}).
To make this notion concrete, we identify four key dimensions that characterize human-centered coding agents---task alignment, steerability, verification, and adaptability---and formalize them with measurable definitions.
We choose these dimensions as interaction primitives that together span the agent task-solving cycle (\autoref{fig:what-users-want}): from interpreting user intent, executing actions under human control, exposing outputs for human assessment, to improving behaviors across repeated use. 
Although they may not constitute an exhaustive list of desirable system properties, these foundational capabilities could support higher-level concepts such as collaboration (\S\ref{sec:what-users-want}).

To address these challenges, we outline a few research directions, including building scalable models of human users, enabling more efficient and task-aware verification, defining measurable interaction-centric evaluation signals, and exploring applications for AI coding agents beyond traditional software engineering.
Together, these efforts call for new infrastructure, benchmarks, and evaluation settings that incorporate human interaction as an essential component, rather than treating it as an external afterthought (\S\ref{sec:closing-the-gap}).

While we argue that human-centered design is essential for coding agents, this view is not without contention. We engage with common counter-arguments and clarify why these perspectives do not eliminate the need for human-centered agent research (\S\ref{sec:alternative-views}).
Collectively, they point toward a research agenda centered not on replacing human developers, but on building AI systems that meaningfully augment us.
\section{AI for Code Today}
\label{sec:ai4c-today}

Recent coding agent research has largely converged on a single objective: \textit{increasing agent autonomy}, typically measured by an agent's ability to complete standalone software engineering tasks \citep{verified,livecodebench}.
Under this framing, ``better'' agents are those that can handle harder benchmarks \citep{deng2025swebenchproaiagents,merrill2026terminalbenchbenchmarkingagentshard}, execute longer horizons \citep{zhao2024commit0librarygenerationscratch}, and achieve higher end-to-end success rates.
This autonomy-centric perspective has shaped several dominant research directions.

First, substantial effort has gone into \textit{constructing more difficult benchmarks}, which demands understanding across languages \citep{swebench-multilingual} and modalities \citep{swebench-multimodal}, execution in longer horizons \citep{li2024promptinglargelanguagemodels,zhao2024commit0librarygenerationscratch} and more complex environments \citep{mlebench,yang2025codeclashbenchmarkinggoalorientedsoftware}. While these benchmarks are all designed to challenge agents with increasingly complex engineering tasks, they may lead the community to overweigh tasks at the tail-end of the difficulty spectrum.

Second, to solve these increasingly difficult problems, researchers have focused on building \textit{sophisticated agent frameworks} that scaffold LMs with crafted pipelines \citep{agentless} and tools \citep{swe-agent,openhands}. Despite differences in architectures, these frameworks share a common goal of pushing agents toward fully autonomous engineering, often limiting human control, which can ultimately constrain their practical utility.

Third, the field has invested in specialized environments and data curation recipes to enable \textit{scalable training with effective verification}. For instance, SWE-Gym established a demonstration-based training formula \citep{swe-gym}, SWE-smith and R2E-Gym automated environment scaling and verification~\citep{swebench-multilingual,r2e-gym}, and later work diversifies task coverage~\citep{sonwane2025bugpilotcomplexbuggeneration,zhu2025trainingversatilecodingagents}. 
These training 
reinforce the autonomy-centric loop, which continuously synthesizes more difficult tasks and trains agents to improve on them.

Although these research have drastically expanded agent capabilities, the assumption that continuing these efforts yields proportional practical gains remains largely unexamined. 
This gap stems from a mismatch between benchmarks optimized for autonomous behavior and real-world settings that inevitably involve human interaction.
If we continue to prioritize isolated autonomy as the primary axis of progress, research risks overlooking how coding agents are actually used: through communication with humans, oversight and intervention from humans, and adaptation alongside humans.
As a result, the field may be optimizing in a direction of diminishing returns, pursuing ever-harder tasks while marginal utility gains shrink, when the gradient toward practical impact points elsewhere: human-agent interaction.

\section{What Users Want}
\label{sec:what-users-want}
Practical deployment of coding agents reveals bottlenecks that differ from those emphasized in automation-oriented agent development.
Industry reports and developer surveys from major coding tool providers consistently highlight how users are often less limited by agents' code correctness than by challenges in communicating with the agent, consuming and verifying its outputs, steering its behavior, and the agent adapting to an evolving codebase and user base.

These practical concerns motivate a shift toward human-centered coding systems design.
We articulate four key pillars for building effective human-centered coding agents.
In addition to qualitative characterization, we formalize these pillars with computable metrics that enable systematic evaluation and optimization.
Our goal is not merely conceptual framing, but to provide actionable targets that can be reported in experiments and optimized by future systems.

\subsection{Task Alignment}
\label{sec:grounding}
\textbf{Definition.}
Task alignment refers to the process by which humans and agents establish and maintain a shared task understanding through mutual modeling~\citep{clark1991grounding}.
In AI-assisted coding, users must externalize intent through NL instructions, relevant contexts, or constraints; while agents must infer latent goals, resolve under-specification, surface assumptions, and signal uncertainty.
Effective alignment is therefore not a single capability, but a coordinated behavior: dynamically integrating interpretation, clarification, and revision as the task unfolds.

\textbf{Formulation.}
Given an NL instruction $q_H$ from user $H$, the agent $C$ forms an internal task specification $z_C = \mathrm{interpret}_C(q_H | \mathcal{T}) \in \mathcal{Z}$ situated in environment $\mathcal{T}$ with task contexts, where $\mathcal{Z}$ denotes a structured intent space.
Task alignment measures distance between the user-intended specification $z_H$ and the agent's inferred $z_C$, 
\[
G(H, C; q_H) = \mathrm{sim}_Z(z_H, z_C)
\] 
where $z_H$ can come from human annotation, and $\mathrm{sim}$ measures task similarity (cosine similarity, lexical overlap).

\textbf{Motivation.}
Task alignment is key to a smooth user experience for coding agents.
Industry has recognized this, producing best-practice guides on prompt engineering, instruction files, and structured frameworks for agent interaction~\citep{github2025wrap,cursor2026bestpractices}.
Studies confirm the stakes: developers who communicate more effectively with agents see measurably better outcomes~\citep{sarkar2025ai}.
When communication fails, the consequences compound.
Incorrect assumptions, wasted computation, and repeated correction cycles degrade both productivity and trust~\citep{mozannar2024reading}.
Today, users largely adapt to agents, learning to phrase requests in ways they understand.
A significant pain point is the ``clarification spiral.''
Given a task, models launch into implementation without inferring unstated constraints and make incorrect assumptions.
Users then reject results and relay additional asks, only to watch new misunderstandings emerge.
What we ultimately want that remains unfulfilled is the reverse: agents that properly model users and interpret tasks.

\textbf{Research Gap.}
Coding dialogue is fundamentally more demanding than traditional task-oriented dialogue~\citep{chi2025edit}.
Whereas a basic request like hotel booking simply requires satisfying a fixed schema of options, programming tasks frequently necessitate diverse communication acts~\citep{ross2023programmer}.

Despite widespread use, open conversation data between humans and AI coding systems remains starkly lacking.
Datasets like WildChat and OpenAssistant capture general dialogue patterns~\citep{kopf2023openassistant,zhao2024wildchat}, but developer dialogue is heavily grounded in external artifacts (code, error logs, documentation) and shaped by tacit expectations around style, architecture, and implementation conventions.
Today, benchmarks sidestep the study of such intricacies entirely: pass@k collapses all signal about communicative quality into a single bit, discarding everything about how mutual understanding was (or was not) achieved.

Without open coding conversation data or widely adopted open-source tooling, systematic and quantitative evaluation of grounding quality remains prohibitively out of reach~\citep{vijayvargiya2025interactiveagentsovercomeambiguity}.
Communication is instead assessed through anecdotal reports of user frustration~\citep{jiang2022discovering,liu2023wants}, leaving us without benchmarks to reliably track progress over time.

\subsection{Steerability}
\label{sec:steerability}

\textbf{Definition.} 
Steerability concerns an agent's ability to expose and respond to human control signals throughout task execution. Rather than optimizing solely for uninterrupted autonomy~\citep{horvitz1999principles}, steerable agents must recognize meaningful decision points in a task, operate at appropriate levels of abstraction, and structure execution to expose decision boundaries at which human intervention can shape downstream behavior.
This enables users to redirect subgoals and adjust execution strategies, while preserving agent autonomy over low-level actions.

\textbf{Formulation.}
Given a user task $q_H$, the agent executes an action trajectory $\tau = (s_1, \cdots, s_N)$ where $s_i$ denotes atomic actions. The agent induces a higher-level segmentation $\mathcal{G}_C(\tau) = \{g_1, \cdots, g_M\}$, where each segment $g_m = (\tau_{k_m:k_{m+1}}, c_m)$ consists of a contiguous sub-trajectory and an associated control point $c_m \in \mathcal{C}$ (e.g., branching choices, confirmation prompts).
To measure agent's responsiveness to human control, let user apply an intervention $u_m$ at the control point $c_m$, yielding a modified trajectory $\tau' = \mathrm{exec}_C(\tau, u_m)$. 
Comparing to a reference trajectory $\tau^*_{u_m}$ reflecting the intended behavioral change, we define response steerability as 
\[
S(H, C) = \mathbb{E}_{(m, u_m)}[{\rm sim}_{\tau}(\tau', \tau^*_{u_m})]
\]
While steerability is fundamentally defined by how agent behavior responds to human interventions, we could additionally measure structural alignment of exposed control points as a diagnostic signal.
Let $\mathcal{G}^*$ denotes a reference segmentation derived from expert demonstration, we define structural steerability as $S_\text{struct}(C) = \mathrm{sim}(\mathcal{G}_C(\tau), \mathcal{G}^*)$, where $\mathrm{sim}$ measures alignment of both segmentation boundaries and control point content.

\textbf{Motivation.}
\vccomment{it's not totally clear to me the diff between alignment and steerability: if you were totally aligned on the task, then not much need to steer outputs?}
Users of AI coding systems consistently express a need to control when and how it acts throughout a task~\citep{kalliamvakou2025newidentity}.
Depending on context, users may operate at different points along the granularity spectrum~\citep{sapkota2025vibecoding}: at times specifying high-level intent for agents to explore broadly via ``vibe code'' and rapid prototyping ~\citep{kalliamvakou2024secondbrain,cursor2026bestpractices}; and at other times intervening at fine-grained levels to make precise, localized changes.
However, these needs cannot be met by systems that treat autonomy as a single global setting. Instead, practical steerability requires agents to identify meaningful control points in the task-solving process---moments where alternative execution paths, tradeoffs, or commitments arise---and to expose those choices to the user. 
By structuring execution around these control points, agents allow users to direct what happens next, rather than forcing them into a binary accept-or-reject role only after a fully specified plan has already been carried out.
While task alignment governs what the user means, steerability governs what the system does next.

\textbf{Research Gap.}
Steerability remains weakly supported in current coding agent systems: existing work is largely descriptive or peripheral, either documenting developer behavior shifts as agent automation increases \citep{chen2025code} or designing human-in-the-loop co-planning \citep{mozannar2025magenticui}, but neither provides a formal account of how agents should structure execution to preserve user control.
At present, we lack principled approaches for agents to identify and expose meaningful control points to users, leaving systems trapped between a false dichotomy between full agent autonomy and constant human supervision.

\vccomment{small observation: this research gap seems unusually short compared to the others}

\subsection{Verification}
\label{sec:verification}
\textbf{Definition.}
Verification refers to a user's ability to assess if a coding agent's outputs are correct with respect to task requirements. This presupposes that users can meaningfully consume and understand the agent's artifacts, including code outputs, execution traces, and intermediate reasoning \citep{vaithilingam2022expectation}. 
In practice, verification determines whether the outputs satisfy both explicit requirements and implicit constraints~\citep{fakhoury2024llm}.
In other words, verification is about whether agent deliverables expose sufficient structure and evidence for humans to accurately judge correctness.

\textbf{Formulation.}
Given a user instruction $q_H$, the agent produces output $o$, including intermediate artifacts and final output.
A human verifier produces a potentially imperfect judgment $s_H = \mathrm{verify}(o, z_H)$ due to agent output verifiability. 
Let $y^*(o, z_H)$ denote the ground-truth verifiers (correctness, efficiency, etc.), 
we define verifiability as the expected agreement between human and gold judgments: 
\[
V(H,C) = \mathbb{E}_{(o,z_H)}[\mathcal{A}(s_H, y^*(o, z_H))]
\]
where $\mathcal{A}$ denotes an agreement metric such as accuracy. 

\textbf{Motivation.}
Users cannot trust what they cannot verify.
Recent surveys find that of the 84\% of developers now using AI coding tools, nearly half do not trust outputs while two-thirds report that half-baked solutions lead to heavier debugging burdens~\citep{stackoverflow2025survey}.
Importantly, current works establish that a user's trust in coding tools is continually recalibrated through repeated verification \citep{sapkota2025vibecoding}, a process that shifts meaningfully with task stakes and user expertise~\citep{wang2024investigating}.
For instance, while experienced developers prefer to inspect low-level implementation details, novice programmers may rely on higher-level natural language explanations and observable artifacts~\citep{barke2023grounded,gu2024analysts}.
Today, the burden is uneven, with less experienced developers reporting both the highest productivity gains and greatest struggle to review coding agent outputs~\citep{sonar2026}.
Supporting verification across this spectrum is therefore critical for enabling reliable and sustained human–agent collaboration.

\textbf{Research Gap.}
Unit testing is the dominant verification paradigm in current benchmarks given their precision and reproducibility.
However, in practice, it captures only a narrow notion of functional correctness and often shifts the verification burden onto users. 
Humans rely on a much broader range of strategies to establish trust in code and the systems that generate it, such as code inspection, execution tracing, and iterative dialogue~\citep{mozannar2024realhumanevalevaluatinglargelanguage}.
This mismatch suggest that while tests are valuable, they may be insufficient for supporting verification in the wild.

First, tests do not always reduce verification effort.
Developers rarely construct test suites before prompting AI tools~\citep{sonar2026}; instead, tests are generated with code.
When AI is producing both, the tests no longer serve as independent evidence for building trust.
Rather, in addition to code, users must now check if tests capture intent -- a lateral shift, not reduction, in cognitive load.
To be clear, unit tests are valuable when constructed strategically, by encoding requirements once to avoid repeated verification.
But using tests as the default mechanism for any agent output creates work that many interactions do not warrant.

Second, while unit testing has its place, it can be ill-suited for non-technical users and tasks beyond traditional software engineering.
Code is increasingly the action space for general-purpose agents, powering exciting use cases such as data analysis~\citep{gu2024analysts}, interface design~\citep{yuan2025towards}, and tasks on the edge of imagination (e.g., raising a plant\footnote{See the \href{https://autoncorp.com/biodome/}{AutonCorp Biodome Project}, where Claude autonomously manages a tomato plant via IoT sensors.}).
Consequently, outputs are becoming increasingly heterogeneous and context-dependent, making universal pass/fail criteria hard to define~\citep{barr2014oracle}.

Rather than relying solely on test-centric workflows, verification should surface evidence in human-interpretable ways, such as visual previews and interactive summaries.
Verification needs to be designed in ways that reduce user burden and make agent behavior easier to assess, especially for users without deep programming expertise.

\subsection{Adaptability}
\label{sec:adaptability}
\textbf{Definition.} We define adaptability as an AI coding agent's ability to maintain and update itself using accumulated experience, to improve future performance while preserving previously acquired capabilities~\citep{trove}. 
This includes updating persistent memory such as user preferences~\citep{letta} and task specifications~\citep{wang2025agent}, as well as action policies by acquiring and refining reusable skills~\citep{wang2025inducingprogrammaticskillsagentic}.
Beyond storing historic context, highly adaptive agents leverage experience to improve how they perceive the environment and take actions, generalize to related tasks while avoiding undesirable drift.

\textbf{Formulation.}
Let $\tau$ depict a distribution (user task, preference, etc.), on which the agent aggregates task experiences $e \sim \tau$ and updates itself $C^{k} = \mathrm{adapt}(C, \{e\}_1^k)$.
We quantify adaptability as the improvement over task sessions 
\[
A(C^k) = \mathbb{E}_{\tau \sim \mathcal{T}} \big[\mathrm{Perf}(C^k) - \mathrm{Perf}(C^0) \big]
\]
where $\mathrm{Perf}(\cdot)$ denotes a task performance metric.
This formulation captures the agent's ability to improve through experience toward a target distribution.

\textbf{Motivation.} Software engineering is inherently adaptive: Codebases are living artifacts that developers tend to, expand, refactor, and repurpose over time.
Users increasingly expect the same continuity and growth from their AI coding tools, yet a common pain point is the need to re-establish context and re-state execution details every session, often described as ``prompt fatigue''~\citep{axifypromptfatigue2025}.
Early efforts in building extensions that auto-generate documentation to maintain persistent memory, suggest reductions in user cognitive load and more efficient, personalized interactions \citep{cursordynamiccontext2026}.
Yet adaptability extends beyond remembering contexts to upgrading their actions, by iteratively developing their skill repertoire and reusing them effectively.
However, the design of such systems is still in its infancy. Today's agent ``memory'' and ``skills'' are often little more than indexed markdown files \citep{anthropicsskills}, rather than mechanisms for sustained learning, leaving the promise of genuinely adaptive coding agents largely unrealized.

\textbf{Research gap.} 
Today's coding agents are developed and evaluated on isolated, one-off tasks, with no incentive for learning from prior sessions or accumulating user-specific context.
Even when tasks share a repository or creator (e.g., 850 SWE-bench tasks from Django), solving one has no effect on a later task.
Simply designing harder problems that require more turns to solve does not obviate the need for persistent context across sessions, a key to sustained support for human engineers.
After all, beyond functional code, repositories embody collaboration across developers, accumulating a collective of conventions, preferences, and institutional knowledge over time.
Systems that internalize this are most effective at reducing the cognitive burden of repeating context over multiple runs.

When adaptability is studied, it optimizes for agents, not users.
Agent Skill Induction~\citep{wang2025inducingprogrammaticskillsagentic} and Live-SWE-agent~\citep{livesweagent} show that agents can acquire reusable skills and improve task performance across sessions, but the metric remains squarely on task success, with no notion of whether or how a human copilot might benefit.
The gaps come to light when collaboration with a human-in-the-loop extends beyond a single session.
A developer tells an agent to use \texttt{logging} instead of \texttt{print}.
Does the lesson persist to next week, or vanish by the morning?
If contradictory preferences are introduced, can the agent sort out the conflicts and ask for confirmation if needed?
As discussed in \textit{Motivation}, practitioners are already improvising solutions (\texttt{AGENTS.md} file, custom rule sets) with promising results.
But systematically evaluating whether current and future proposals improve multi-session human-agent collaboration remains uncharted.

\subsection{Isn't there more?}

Beyond these four pillars, readers may naturally wonder about other properties, such as safety or proactivity.
As discussed in \S\ref{sec:introduction}, we view additional directions as complementary rather than foundational.
They either emerge as compositions of core dimensions (e.g., safety depends on grounding and verification working together) or reflect system-level design choices, not underlying interaction capabilities.
Find discussions of several such properties in \S\ref{appx:complement_pillars}.

\section{Closing the Gap}
\label{sec:closing-the-gap}

To tackle the gaps from \S\ref{sec:what-users-want}, rather than prescribing isolated fixes per area, we consider: ``What infrastructure missing today blocks progress across multiple dimensions?"

A common bottleneck across all pillars is that our formulations require sampling distributions of human behavior -- intent, judgments, interventions -- that current open-source pipelines do not provide at scale.
Furthermore, we lack task structures that truly require interaction to succeed and demand verification approaches beyond unit tests, particularly in application domains beyond software engineering.

We identify four high-leverage research directions, each targeting a bottleneck that stalls multiple fronts simultaneously.
For each, we discuss how progress is being throttled, then propose potential solutions.

\subsection{Scale human modeling}
Today, researchers face an unfortunate dilemma when evaluating human-AI collaboration: either run expensive human studies that do not scale, or default to fully autonomous benchmarks that sidestep the human entirely.
The machine learning community has largely chosen the latter.
But for genuine progress on human-centered coding agents, there is no shortcut.
The formulations in \S\ref{sec:what-users-want} all require sampling from human distributions: specification ($\mathcal{G}$) necessitates user intent $z_H$, verification ($\mathcal{V}$) depends on human judgments $s_H$, and steerability ($\mathcal{S}$) relies on human intervention $u_m$.

\begin{figure}
    \centering
    \includegraphics[width=0.9\linewidth]{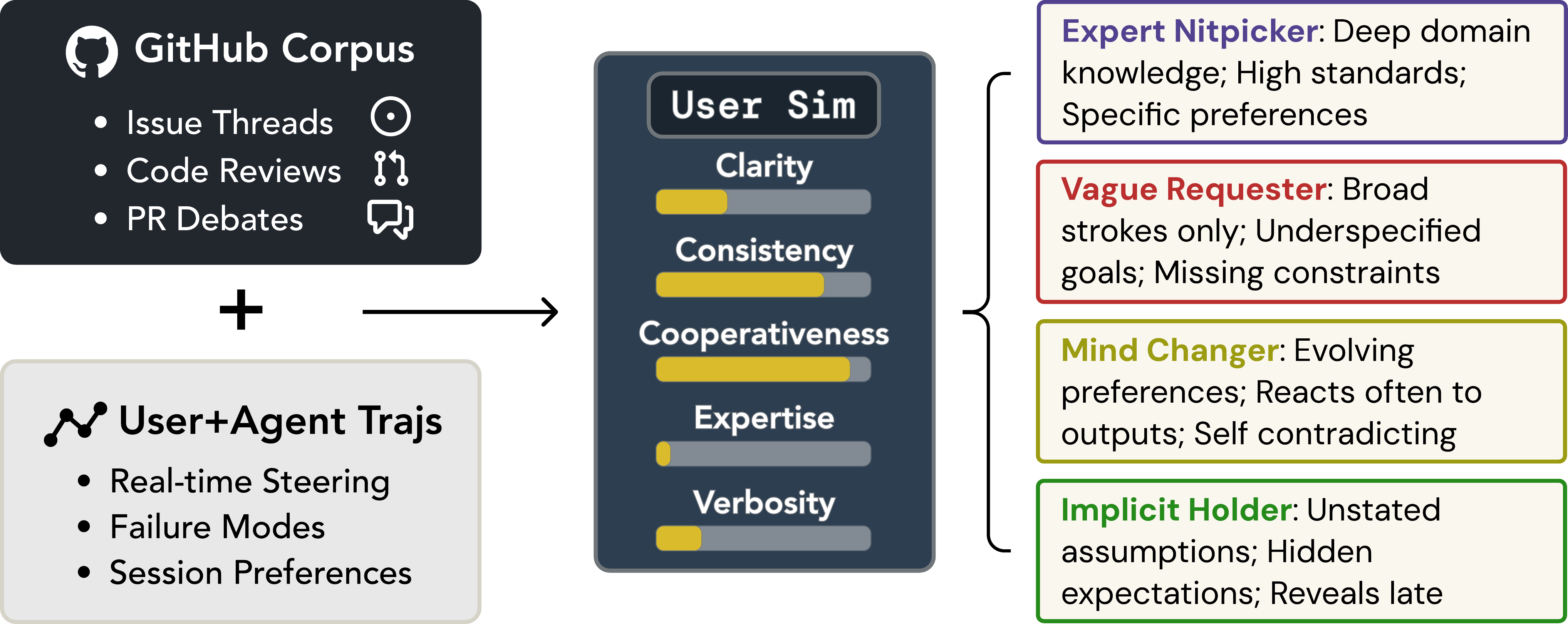}
    \caption{
    \textbf{Generating Diverse Simulated Users.} We envision combining GitHub interaction data with observed human-agent trajectories to parameterize user simulators along dimensions like clarity, consistency, and expertise.
    }
    \label{fig:profile_generator}
\end{figure}

To address this, the first path is to construct executable environments with (simulated) users and establish training pipelines accordingly. This begins with developing \textit{user simulators} that faithfully model humans as active software developers and users~\citep{taubench2024}. 
A key challenge is that current simulators prompted to ``act like a human'' behave homogeneously and overly cooperative~\citep{li2017usersimulatortaskcompletiondialogues}.
More prompting alone is not sufficient:
faithful simulation requires modeling personal preferences, expertise, and realistic failure modes, since real users often hold implicit expectations they may not always verbalize and change their minds mid-task~\citep{naous2025flippingdialoguetrainingevaluating,shi2025impersonaevaluatingindividuallevel}.
One approach is to leverage large-scale developer profiles and interactions available on GitHub (\autoref{fig:profile_generator}), such as pull request histories, even real user-agent interaction data, which together capture user behavioral and preferential patterns to train more realistic interlocutors.

A complementary approach is to build widely adopted user-facing platforms that \textit{collect interaction data as a byproduct of real use}.
For instance, Copilot Arena integrates into IDEs and gathers large-scale preference signals on daily coding tasks that substantially diverge from model rankings from static benchmarks~\citep{chi2025copilotarenaplatformcode}.
We encourage researchers and industry to partner and share anonymized interaction data to enable scalable research.

Ultimately, measuring and improving the fidelity of LM-based simulators remains an open challenge.
How do we evaluate alignment to user intent and behavior, ensure diversity without collapsing to stereotypes, and effectively combine simulated and real interaction signals?

\subsection{Enable efficient oversight}
As discussed in \S\ref{sec:steerability} and~\S\ref{sec:verification}, today's users verify agent outputs in tedious and manual ways.
Proxies of code correctness, such as unit tests, scale poorly in the wild and may not be well-suited to the task or end user's expertise.

Rather than placing the full burden on users, coding systems should proactively support oversight \citep{chen2025needhelpdesigningproactive,sun2025trainingproactivepersonalizedllm,zhao2025codinggenieproactivellmpoweredprogramming,raghavendra2026agentic}.
Recent work has explored generate relevant tests without explicit requests~\citep{chen2022codet,mundler2024swtbench,jain2025testgenevalrealworldunit} and self-checking outputs prior to delivery~\citep{chen2023teaching,ni2023lever}.

\begin{figure}
    \centering
    \includegraphics[width=0.9\linewidth]{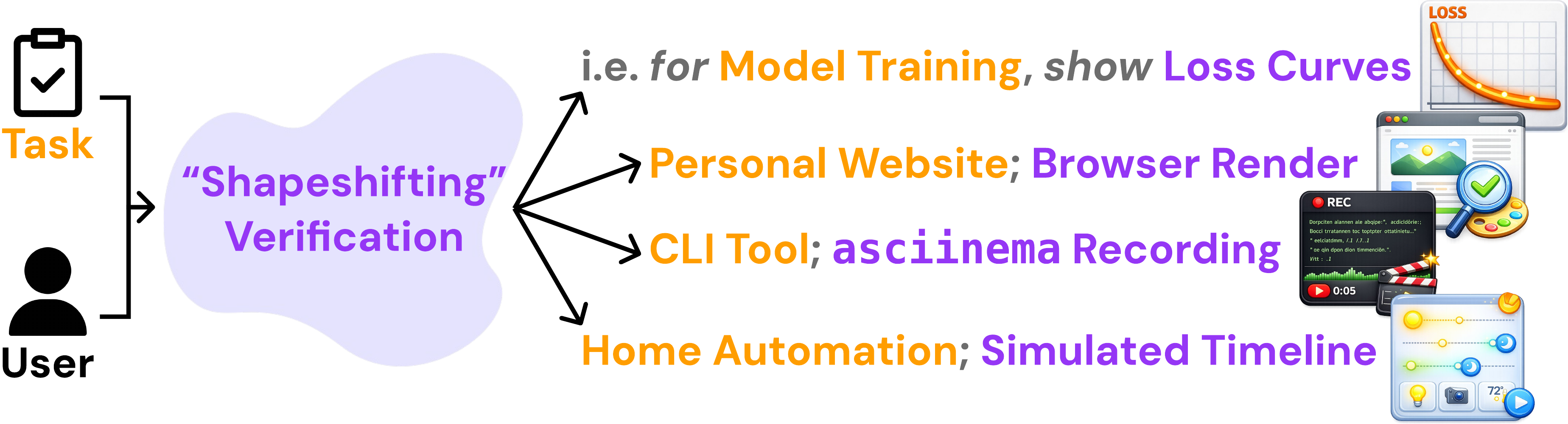}
    \caption{
    \textbf{Context should determine the ``shape" of oversight.} Rather than defaulting to unit tests or code diffs, AI systems should surface artifacts that make validation intuitive.
    }
    \label{fig:shapeshifting_verification}
\end{figure}

Supplanting such traditional methods, we envision verification becoming a dynamic, protean procedure.
Conditioned on task and user, agents should reason about notions of quality and surface appropriate artifacts to make validation tractable, as imagined in Figure~\ref{fig:shapeshifting_verification}.
A data pipeline implementation might warrant a \texttt{.md} file summarizing key steps (e.g., how were missing values handled).
On the other hand, code for a webpage should be presented visually for humans to assess design quality at a glance.

Several open questions remain.
How should we evaluate the quality of verification approaches themselves? 
What communicative cost can users bear before cognitive load degrades their judgment? 
And when should agents self-verify versus surface as artifacts for human review?

\subsection{Define measures for interaction}
The current focus on task resolution leaves interaction quality unmeasured and unoptimized.
\S\ref{sec:what-users-want}'s formulations all involve expectations over human behavior, such as intent ($z_H$), judgments ($s_H$), or interventions ($u_M$).
We hypothesize that operationalizing such latent variables has value.
How many turns does it take to recover $z_H$?
How often is $u_M$ required?
How much human effort does $s_H$ demand?

Fortunately, we need not start from scratch.
Decades of HCI and software engineering research offer rich precedent for defining and operationalizing interaction quality from user studies~\citep{sarkar2022like,sun2022investigatingexplainabilitygenerativeai,sergeyuk2024ide}.
\citet{mozannar2024reading} introduces the CUPS taxonomy, a categorization of programmer-AI interaction into 12 discrete states (e.g., ``Prompt Crafting", ``Editing Recent Suggestions"), which they then use to measure the distribution of time and transitions across these states.
\citet{barke2023grounded} distinguishes acceleration mode (user knows what to do, measures time-to-completion) from exploration mode (user uncertain, measures validation effort and suggestion-browsing depth).
\citet{chen2025can} proposes the PULSE framework to operationalize user satisfaction as a training signal, using models to predict satisfaction from interaction traces and project the effect of agent design choices.
Literature in such fields is ripe with such gems; model trainers and agent builders could port and operationalize them at scale.

Beyond adapting existing frameworks, there is gold to be found in large-scale trajectory analysis.
Cursor's Tab RL work \citep{cursortabrl2025} illustrates this: by mining 400M+ daily accept/reject decisions, they discovered that \textit{when} to suggest matters as much as \textit{what} to suggest.
This insight is invisible to pass-rate benchmarks.
Infrastructure for agent trajectory analysis \citep{bouzenia2025understanding,stringsight_github,meng2025docent} presents an exciting opportunity to rapidly define and validate human-centered interaction metrics like collaboration quality and user satisfaction.

\subsection{Go beyond software engineering}
Coding agents are general agents~\citep{soni2025coding}.
Any task expressible programmatically becomes fair game.
Manage a stock portfolio.
Control a smart home.
Tend a greenhouse.
Plan a wedding.
Compose music.
People are already using coding agents for these and more.
This shift demands that agent design move beyond tools optimized for professional developers to support everyday users.

What makes such domains compelling is that the four pillars arise intrinsically.
Grounding user intent is essential: when an agent managing a portfolio is told ``be more conservative", this means different things from a retiree versus a college graduate.
Orchestrating a smart home foregrounds verification; if ``unlock the front door" or ``turn on the stove" leads to mistakes, the damage can't be undone with a simple \texttt{git revert}.
Steerability is indispensable for planning tasks, where flights get canceled, open slots get booked, and user preferences change.
Consider monitoring a greenhouse.
The right watering schedule depends on this morning's humidity, but optimal planting strategy is best informed by last season's growing cycles.
An agent must digest feedback signals spanning hours to months, deciding which timescales to attend for each decision.
This is adaptability at its purest.

\section{Alternative Views}
\label{sec:alternative-views}

We dedicate the following section to addressing meaningful points of contention to arguments put forth by our position.

\textbf{(a)} \textit{Given the rapid rate of progress, AI will take over coding.
Studying how humans interact and collaborate with AI coding systems is a fleeting need.}

Anthropic CEO Dario Amodei famously stated in March 2025 that in the near future, AI will write 90\% of code~\citep{amodei90percentcoder2025}, a declaration representative of a general sentiment that ongoing improvements in models' coding capabilities will eventually stamp out any need for human effort in software development~\citep{simon1965shape,huangcodeisdead2024}.

We posit that whether or not this future materializes, the study of how humans participate and benefit from an AI-enhanced coding loop will remain relevant.
Even if Dr. Amodei's prediction bears out, humans do not exit the loop.
As agents carry out increasingly complex tasks, more decisions must be made, many of which are personalized, organization-specific, or depend on information only a specific individual possesses.
The locus of effort simply shifts upstream, from implementation to specification and evaluation.
Code, after all, is simply the language we use to communicate with machines -- one that continues to evolve, from assembly to C to Python.
Assuming agents bear more responsibility for writing Python, developers may instead convey intent with natural language or perhaps a new formalism balancing human expressiveness with programmatic precision.
But the task of coding, fundamentally how we communicate with machines, does not vanish; it transforms.

Present-day trends already reflect this.
Concurrent with the emergence of more end-to-end coding agents~\citep{claudecode,openaicodex}, tools designed to aid developers, not replace them, have proliferated: for reviewing~\citep{graphite,greptile}, editing front-ends~\citep{cursorvisualeditor2025}, completion~\citep{githubcopilot}, and debugging~\citep{levin2024chatdbg,databricksdebug2025}.
The spectrum of tools coming to market suggests that users not only prefer copilots in some cases, but also desire different interfaces for different tasks.
Both in-IDE auto-complete and a visual editor can be used to modify webpage elements, but which a developer reaches for depends on the task, their expertise, and preferences.
Beyond tools, the outcomes also embed personal taste.
Should a button be red or blue?
Styled with inline CSS or reusable classes?
Accessed via a modal or new page?

\textbf{(b)} \textit{Human interaction and evaluation are too costly to scale.}

Human evaluation for natural language generation systems is challenging~\citep{celikyilmaz2020evaluation}.
Studies are expensive and time-consuming, particularly for tasks requiring heavy domain expertise like coding~\citep{howcroft-etal-2020-twenty}.
A lack of standards around how to design, execute, and report human evaluations can easily lead to irreproducible results~\citep{van2021human}.

These concerns, while valid, can be overcome.
As covered in \S\ref{sec:closing-the-gap}, just as the NLP community developed scalable proxies of human feedback to improve instruction following and alignment (e.g., reward models, LM-as-judge~\citep{dubois2023alpacafarm,zheng2023judging}), metrics such as cyclomatic complexity~\citep{mccabe1976complexity}, cognitive complexity~\citep{campbell2018cognitive}, maintainability index~\citep{oman1992metrics} and readability~\citep{buse2008metric,buse2009learning} are computable and tangibly alleviate maintenance burden.
More ambitiously, recent works have showcased the viability of benchmarks with user simulators~\citep{taubench2024} and preference collection through organic usage~\citep{chiang2024chatbot} as paths towards scalable human-centered evaluation.
The challenge of standardizing such nascent approaches is precisely why now is the time to act~\citep{shao2025futureworkaiagents}.

\textbf{(c)} \textit{Capabilities first. Human-centered concerns are product problems that model builders shouldn't get distracted by.}

The Bitter Lesson argues that general methods leveraging computation ultimately supersede manually crafted approaches based on human knowledge of a domain~\citep{sutton2019bitter}.
If anything, the current market reflects this division of labor.
Researchers optimize capabilities, then product builders graft the appropriate user interfaces downstream.

Our position is not anti-scaling.
Rather, we agree, with the important caveat that what we're actually scaling toward matters.
If we continue optimizing solely for autonomous coding agents, we will produce just that.
Better collaborators will not emerge for free~\citep{shao2025collaborativegymframeworkenabling,shen2025completionneqcollaborationscaling,weston2025ai,wu2025collabllmpassiverespondersactive,zhou2025sweetrltrainingmultiturnllm}.
The leap from GPT-3 to ChatGPT illustrates this point.
Incorporating human preferences into the training objective itself transformed a research artifact into a gripping conversationalist~\citep{ouyang2022training}.
Deferring human-centered concerns to downstream product work risks entrenching interaction patterns that are difficult to undo.

\section{Conclusion}
We contend that the omission of consideration for humans in AI coding agent research threatens to undermine the very utility these systems seek to provide.
Left unchecked, the gap between AI coding agent research and real-world utility may very well widen.
The research community must decide whether to optimize for leaderboards or the people who actually use these systems as well.
Otherwise, we run the risk of building ever-more-capable coding agents that fewer people know how to wield.

\section*{Impact Statements}

We highlight two potential impacts of this position paper on the research community and broader society.

First, our work aims to point out a potential symbiosis between HCI research and agent development in the ML community. While HCI studies how people interact with AI coding systems, much agent research continues to prioritize autonomous task completion. By introducing measurable human-centered objectives, we provide a shared framework that can foster closer integration across ML, HCI, software engineering, and related fields.

Second, our position reframes automation in software work from maximizing autonomy to supporting meaningful human interaction. Rather than replacing programmers, human-centered coding agents can amplify human judgment and initiative while remaining controllable and interpretable. By advocating for accessible interaction and verification mechanisms, this approach also broadens access to programming capabilities for non-experts, enabling wider participation in AI-powered tools.
More in \S\ref{appx:on-the-horizon}.
\section*{Acknowledgments}
We would like to thank members of Stanford NLP, the SALT Lab, and the Language Technologies Institute at CMU for their helpful discussions and insightful feedback on the project.
We would also like to thank Mike Merrill, Ofir Press, Alexander Wettig,  Wenting Zhao, and many attendees of the Deep Learning for Code (DL4C) workshop for providing thoughts and opinions that were formative to the shape of the arguments we put forth.
Zora Zhiruo Wang is supported by Google PhD Fellowship.
This work was supported by an HAI grant, DSO lab, Open Philanthropy, Schmidt Sciences, and a grant under the NSF CAREER IIS-2247357 and ONR N00014-24-1-2532. 

\bibliography{main}

\newpage
\appendix

\section{Metrics}

In this section, we review our methodology for computing ``human-centered'' metrics related to code quality that we used to create the graphs in the main paper.
We also include additional plots and analysis discussing further findings in our investigation of how well code meets standards beyond functional correctness.
\subsection{Patch bloat}
\label{sec:appdx:metrics:conciseness}
Our patch bloat metrics quantify the size of the changes that the LM applied (\emph{submitted patch}) relative to the human solution (\emph{gold patch}).
Consistently large patches are not only a problem for code review and \emph{verifiability}, but can also point to over-engineered or overly complex solutions (again hampering verifiability) or \emph{task grounding} issues.

We compute these metrics for models evaluated on the main SWE-bench leaderboard (SWE-bench Verified, using mini-swe-agent). To avoid confounding effects, we restrict analysis to successfully resolved task instances.

First, we clean both the submitted patch and the gold patch by removing
%
\begin{enumerate}[label=(\roman*)]
    \item all newly added files (for example, this removes reproduction scripts and other auxiliary files from the submitted patch),
    \item all non-Python files (SWE-bench instances are all from Python repositories; this removes pure documentation changes and any other artifacts), and
    \item changes to all test files (we want to focus on the quality of the implementation and remove work on tests as a confounding factor).
\end{enumerate}
We then calculate the \emph{bloat ratio} as the ratio of character lengths between the submitted and gold patch.
We show results for two metrics
\begin{enumerate}
    \item \emph{Average bloat ratio:} The average of the bloat ratio across all resolved task instances.
    \item \emph{Bloated patch fraction:} The fraction of resolved task instances where the bloat ratio exceeds 1.5.
\end{enumerate}

\begin{figure}
    \includegraphics[width=\textwidth]{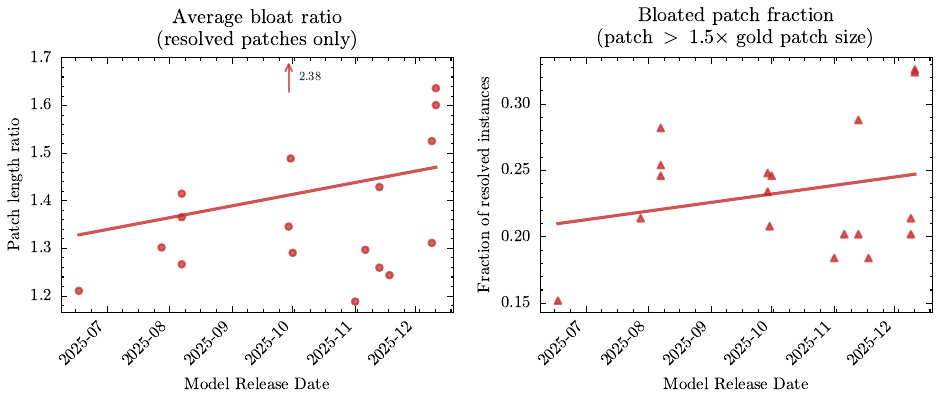}
    \caption{LM generated patch sizes are on the rise.}
    \label{fig:appdx:patch_size_vs_time}
\end{figure}

As observed already in Figure~\ref{fig:teaser}, not only do LMs have a consistently high average bloat ratio, but it also is largely uncorrelated with the task solving ability of the models as measured by the task resolution rate.
Figure~\ref{fig:appdx:patch_size_vs_time} further shows that there is also a positive trend with model release date in both average bloat ratio and bloated patch fraction.
All models show a bloated patch fraction of more than 15\%, demonstrating that the high average bloat ratio is not only an artifact of outliers.
\subsection{Further insights on patch bloat}
\begin{figure}
    \centering
    \includegraphics[width=\linewidth]{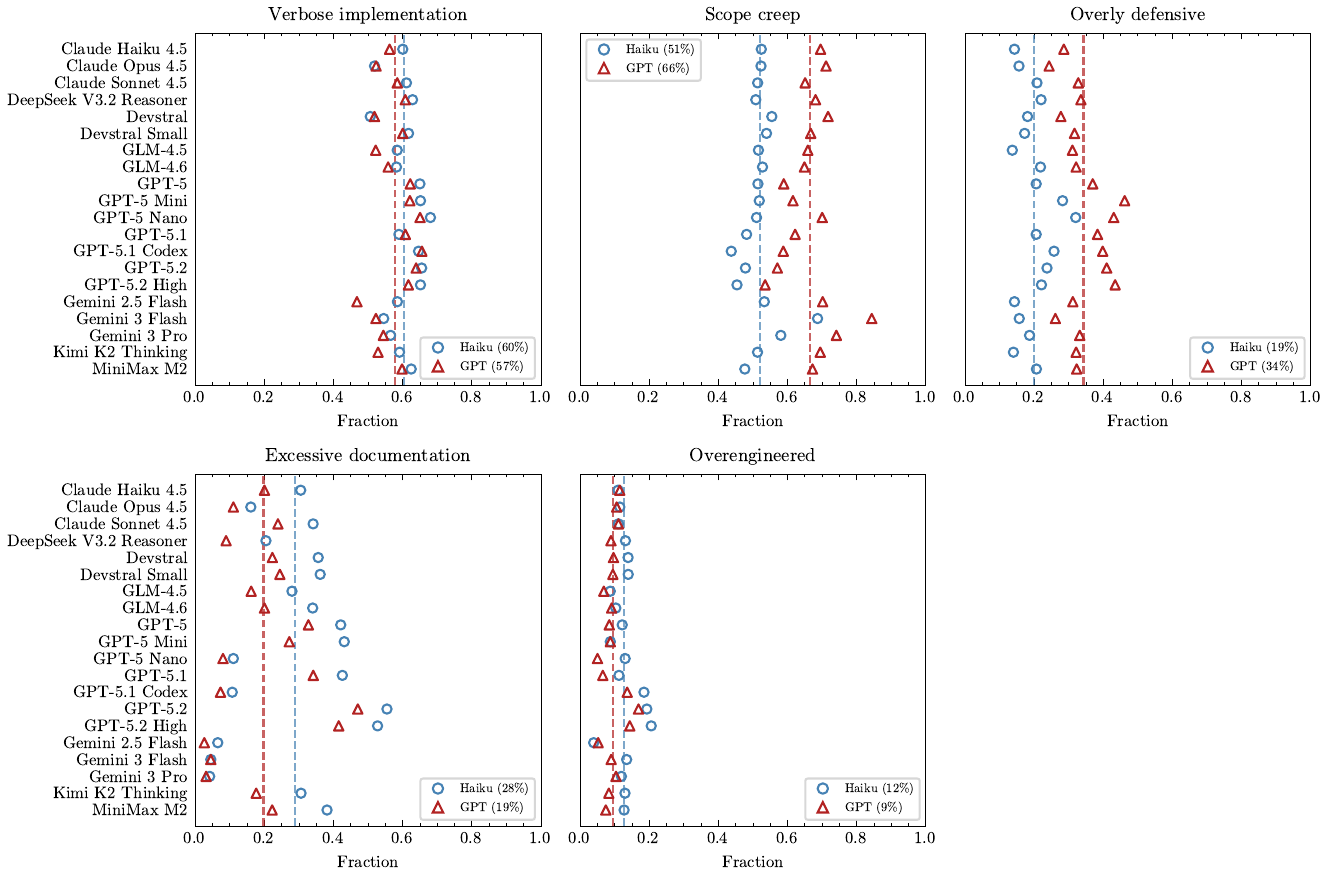}
    \caption{Annotation of factors that cause LM submitted patches to be longer than the human gold solution. Fraction refers to the fraction of submitted patches that resolve the task instance but are longer than their gold patch counterpart. The model average is shown as dashed lines with its numerical values quoted in the figure legends.}
    \label{fig:appdx:details_bloat}
\end{figure}

We annotate all submitted patches that are longer than their gold patch counterpart (i.e., that contribute to an average bloat ratio $>1$) using GPT-5 mini and Claude Haiku 4.5 as a judge to identify the model behaviors that drive patch bloat.
The results are shown in Figure~\ref{fig:appdx:details_bloat}.
Both models flag \emph{verbose implementation} (e.g., unnecessary assignment of intermediate variables) as one of the leading causes for the longer patches, affecting around 60\% of bloated resolved patches.
Next most prevalent is \emph{scope creep} (50--65\%) and \emph{overly defensive} code (20--30\%).
Excessive documentation and generally over-engineered solutions are flagged for around 20--30\% and 10\% of bloating.

The following system prompt was used for annotations:
\begin{lstlisting}[style=prompt]
You are an expert software engineer analyzing why a submitted patch is LONGER than a gold (reference) patch.
Both patches SUCCESSFULLY solve the same problem, but the submitted patch has more lines of code.
Your task is to identify the reasons for this extra length.

The gold patch is assumed to be the minimal, optimal solution.

You are analyzing a submitted patch that is LONGER than the gold patch. Both patches solve the same problem.
Your task is to identify WHY the submitted patch has more lines of code.

Use ONLY these 5 categories:

## overly_defensive

The submitted patch adds defensive code that the gold patch does not include:
- Try/except blocks or error handling not present in the gold patch
- Type checks, None checks, or input validation not present in the gold patch
- Broader exception catching than necessary
- Explicit error returns or fallback paths that aren't needed
- Defensive API options (e.g., errors='ignore') not in the gold patch

Use this category when the extra code is about "being safe" or handling edge cases 
that the gold patch trusts won't occur.

## scope_creep

The submitted patch makes changes beyond what's needed to fix the issue:
- Modifying behavior in ways unrelated to the fix
- Adding features or functionality not requested in the problem statement
- Reformatting existing code (whitespace, quotes, import ordering, line breaks)
- Modifying dependencies and imports
- Touching files or functions that the gold patch doesn't touch
- Removing or adding comments unrelated to the fix

Use this category when the extra code comes from doing MORE than what was asked,
not from doing the same thing in a longer way.

## excessive_documentation

The submitted patch adds comments or documentation that the gold patch does not:
- Inline comments explaining obvious code
- Multi-line comment blocks describing the change
- Docstring additions or modifications not in the gold patch
- Comments that describe "what changed" rather than "why" (temporal comments)
- Redundant documentation of self-explanatory logic

Use this category when the extra lines are comments/docs, not executable code.

## verbose_implementation

The submitted patch implements the same fix but with more code than necessary:
- Writing explicit loops instead of comprehensions or built-in functions
- Duplicating logic that could be factored out or reused
- Not using existing utility functions, APIs, or methods that the gold patch uses
- Using multiple statements where one would suffice
- Overly explicit variable assignments or intermediate steps
- Longer conditional chains that could be simplified

Use this category when the submitted patch does the SAME thing as the gold patch
but uses more lines to express it. The logic is equivalent, just wordier.

## overengineered

The submitted patch introduces unnecessary abstraction or architectural complexity:
- Creating new classes, functions, or modules that the gold patch doesn't need
- Adding configuration options or parameters for flexibility that isn't required
- Implementing generic solutions when a specific fix would suffice
- Adding layers of indirection (wrappers, decorators, factories) not in the gold patch
- Building infrastructure for future extensibility that isn't asked for

Use this category when the extra code comes from ARCHITECTURAL overhead - new 
abstractions, indirection, or generalization beyond what the problem requires.

For each reason you identify, provide:
- category: One of the exact category names listed above
- reason: A brief explanation of this specific instance (1-2 sentences)

You may assign multiple categories if the extra length has multiple causes.
You MUST return at least one category.
\end{lstlisting}
Together with the following user prompt:
\begin{lstlisting}[style=prompt]
<problem_statement>
{{ problem_statement }}
</problem_statement>

<gold_patch>
{{ gold_patch }}
</gold_patch>

<submitted_patch>
{{ submitted_patch }}
</submitted_patch>

The submitted patch is LONGER than the gold patch. Identify the categories that explain this extra length.
You MUST return at least one category.
\end{lstlisting}
\subsection{Functional differences}
SWE-bench task instances are scored as resolved (1) or unresolved (0) based on whether all unit tests (including some that were not visible to the agent during inference time) are passing.
However, tests rarely cover all behavior details\footnote{In particular, because there is a balance to be struck with tests not getting overly specific with respect to the specifications in the issue text}.
This means that even resolved instances can have functional discrepancies with the gold patch.
This does not necessarily mean that they are incorrect solutions to the problem statement, but might also be due to ambiguities or degrees of freedom of the problem statement and implementation decisions that nonetheless have significant downstream impact.
Seeing functional discrepancies even in patches that pass unit tests therefore points to the need for dialogue-based \emph{task grounding} (to resolve ambiguities and obvious missing specifications), \emph{steerability} (to keep the user in the loop for implementation decisions that are not obvious at the beginning of a trajectory) and good \emph{verification}. 

To quantify this amount of functionally discrepant patches, we perform an annotation similar to \S\ref{sec:appdx:metrics:conciseness}.
We perform the same cleaning of patches, and then use Claude Haiku 4.5 and GPT-5 mini as a judge to annotate the tuple of problem statement, submitted patch, and gold patch.
The LM responds with structured output, returning a list of possible issue categories together with reasoning about why they apply.

\begin{figure}
    \includegraphics[width=\textwidth]{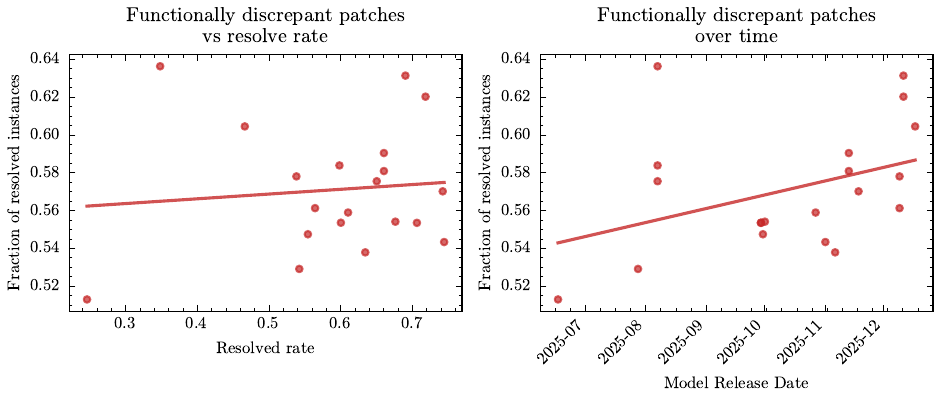}
    \caption{Functional discrepancies in resolved patches}
    \label{fig:appdx:functional_discrepancies}
\end{figure}

To adopt the most conservative stance, we only consider a submitted patch as functionally discrepant if both models flag at least one functional discrepancy.
The results are shown in Figure~\ref{fig:appdx:functional_discrepancies}.
We observe that:
\begin{enumerate}
    \item The discrepancy is higher than 50\% for all models.
    \item Resolve rate is weakly correlated with an increase in functional discrepancies.
    \item Functional discrepancies are more prevalent in patches submitted by more recently released models.
\end{enumerate}

\begin{figure}
    \centering
    \includegraphics[width=\textwidth]{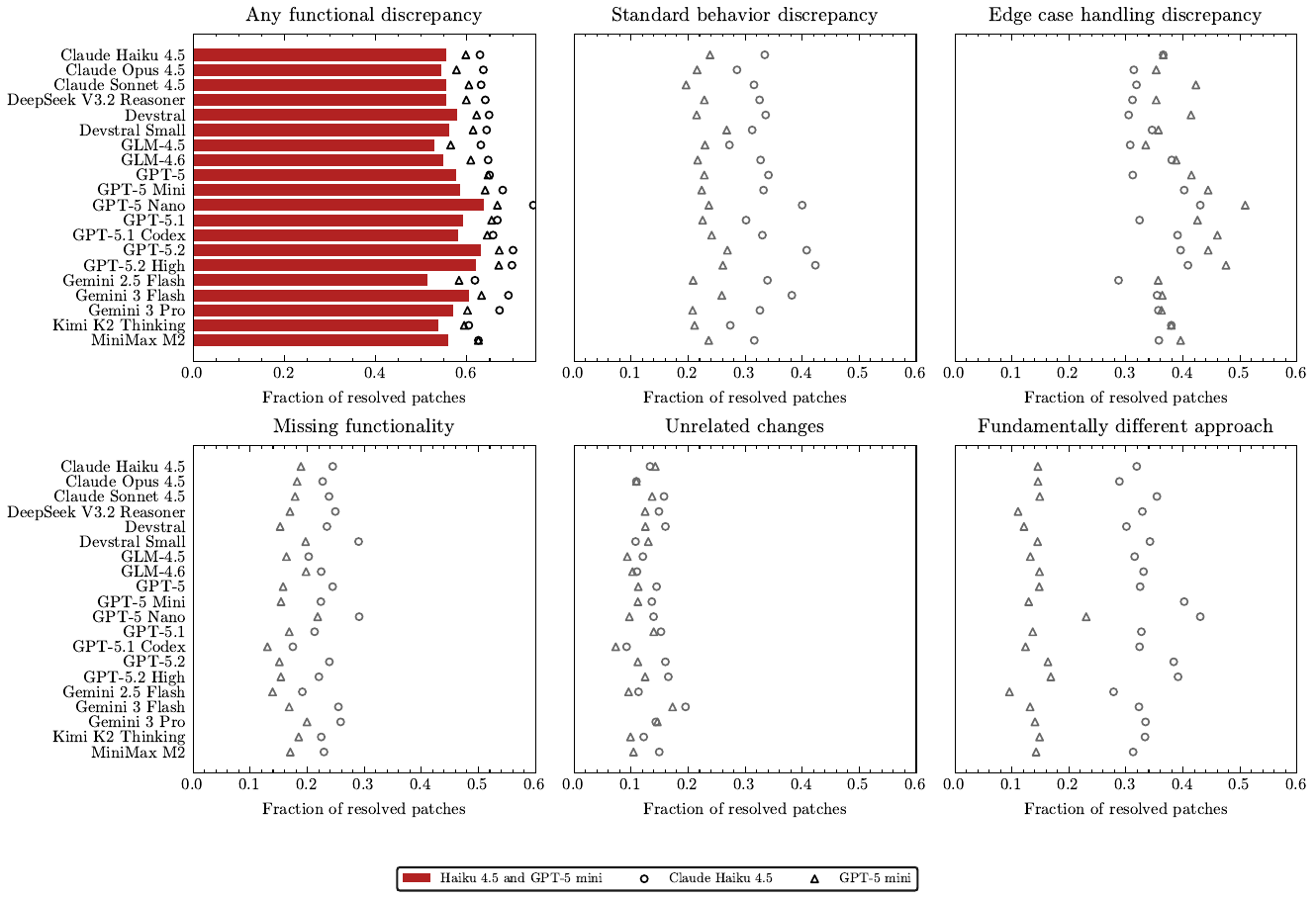}    
\caption{Details on the result of annotating the submitted patches for functional discrepancies with Claude Haiku 4.5 and GPT-5 mini. The top left bars show the same numbers as Figure~\ref{fig:appdx:functional_discrepancies}, conservatively calculated as the fraction of patches that are flagged by both models. The remaining charts show annotation results for specific categories. Because both models make somewhat different choices in how they categorize the same functional discrepancies, there is somewhat less agreement per category than for the overall assessment.}
\label{fig:appdx:details_discrepancies}
\end{figure}

Details on the results of both models are shown in Figure~\ref{fig:appdx:details_discrepancies}.
The number of instances that are flagged by either model are relatively consistent: Across all experiments, Claude Haiku 4.5 flags 66\% of resolved patches as inconsistent, GPT-5 mini 62\%.
The fraction of patches flagged by both models at the same time is 57\% (this corresponds to the numbers shown in Figure~\ref{fig:appdx:functional_discrepancies} and the red bars in Figure~\ref{fig:appdx:details_discrepancies}).
Because the categories of the discrepancies cannot be separated clearly, there is some disagreement between the models when considering individual categories (for example, whether a discrepancy affects \enquote{standard behavior} or \enquote{edge case handling}).
Standard and edge case behavior each affect around one third of patches.
Around 20\% of patches miss functionality that is included in the gold patch while more than 10\% of patches include unrelated changes that aren't included in the gold patch.

For the annotation we use the following system prompt:
\begin{lstlisting}[style=prompt]
You are an expert software engineer analyzing patches/diffs for code changes.
Your task is to compare a submitted patch (<submitted_patch>) with a gold (reference) patch (<gold_patch>).
The gold patch is the reference solution assumed to be correct.
Your job is to identify and categorize any functional discrepancies in the submitted patch relative to the gold patch.

## Definition of "Functional"

A "functional" change affects the observable behavior of the code, including 

- return values, 
- exceptions raised, 
- side effects, 
- I/O, 
- externally-visible state changes.

Treat the gold patch as the reference even if you personally disagree with it.

The following are NOT functional changes:

- Logging or print statements
- Wording of messages that are not clearly specified in the problem statement
- Documentation or comments
- Code formatting or style
- Import ordering
- Variable/function naming (unless it changes a public API that callers rely on)
- Implementation details that produce identical behavior

Message text IS a functional discrepancy only if the problem statement explicitly specifies the exact message content or a strict message format.

## Categories

Use ONLY these categories:

## standard_behavior

The submitted patch produces different behavior than the gold patch for standard, typical inputs implied by the problem statement.
This excludes differences in message wording, comments, or non-functional aspects.

## edge_cases_handling

The submitted patch handles edge cases or error conditions differently than the gold patch.
Edge cases include: null/None values, empty collections, zero values, boundary values, or error paths.
Use this category when the difference is limited to exceptional/boundary conditions; if the difference affects typical inputs, use `standard_behavior` instead.

Examples:
- Additional error return paths not in the gold patch
- Adding options like `error='ignore'` not present in the gold patch
- Catching exceptions and suppressing them (e.g., replacing with print statements) instead of propagating
- Returning different values for boundary inputs

NOT a functional discrepancy:

- Re-raising caught exceptions with the same or equivalent error type
- Different wording of error messages

## missing_functionality

The submitted patch omits functionality that is present in the gold patch.
The gold patch implements something that the submitted patch does not implement at all.
This is different from different_behavior - here the submitted patch simply lacks the functionality entirely.
Use this when the gold introduces a new behavior/branch/check/output and the submitted does not implement it in any form.

## unrelated_changes

The submitted patch makes functional changes unrelated to the problem statement that are also absent from the gold patch.
This includes new features or enhancements beyond the scope of the problem.
This excludes non-functional changes (documentation, comments, formatting, imports).

## fundamentally_different_approach

The submitted patch solves the problem using a fundamentally different approach than the gold patch.
Example: Modifying different (non-private) functions to achieve the solution in such a way that 
there are non-trivial functional discrepancies as a result.

Note: This category describes the structural approach. 
If a fundamentally different approach also causes specific behavioral discrepancies 
(e.g., edge case handling differences), report BOTH:
- One `fundamentally_different_approach` item for the structural difference
- Separate items for each distinct behavioral discrepancy (e.g., `edge_cases_handling`)

## Output Format

You MUST return a single JSON object with exactly this shape:
- reasons: a list of objects, each having:
- category: one of the exact category paths listed above
- reason: a brief explanation (1-2 sentences) that names the triggering condition/input and the behavioral difference (gold vs submitted)

Return a separate item for each distinct functional discrepancy. You may use the same category multiple times.

Example: If the submitted patch 

(1) modifies different functions than the gold patch, 
(2) handles None values differently, 
(3) handles empty lists differently, and 
(4) adds an unrelated feature, 

return four items in `reasons`:

- `fundamentally_different_approach` (for #1)
- `edge_cases_handling` (for #2)
- `edge_cases_handling` (for #3)
- `unrelated_changes` (for #4)

If both patches are functionally identical, return:
{"reasons": []}
\end{lstlisting}
The user prompt is formatted with the problem statement, submitted patch and gold patch as follows:
\begin{lstlisting}[style=prompt]
<problem_statement>
<problem_statement>
{{ problem_statement }}
</problem_statement>

<gold_patch>
{{ gold_patch }}
</gold_patch>

<submitted_patch>
{{ submitted_patch }}
</submitted_patch>

Categorize any functional discrepancies in the submitted patch compared to the gold patch.
If there are no functional discrepancies, return {"reasons": []}.
</submitted_patch>

Categorize any discrepancies (if present) of the submitted patch compared to the gold patch.
It's ok to return an empty list if there are none.
\end{lstlisting}
The following structural output format is enforced:
\begin{lstlisting}[style=python]
CategoryType = Literal[
  "standard_behavior",
  "edge_cases_handling",
  "missing_functionality",
  "unrelated_changes",
  "fundamentally_different_approach",
]

class Reason(BaseModel):
    category: CategoryType
    reason: str

class PatchAnalysis(BaseModel):
    reasons: list[Reason]
\end{lstlisting}
\newpage
\section{Extended Discussion}

\subsection{Complementary Considerations}
\label{appx:complement_pillars}

In addition to the four core dimensions discussed in \S\ref{sec:what-users-want}, we recognize several complementary properties influencing the deployment of human-centered coding agents, though we argue they are best understood as emerging mechanisms rather than primary interaction primitives.
Below, we highlight several considerations that came up during the process of putting together our position.
Per area, we provide our brief definition, explain our reasoning for viewing it as complementary, and clarify how it relates to or emerges from the four core dimensions.

\textbf{Proactivity.}
Proactive behavior refers to an agent's ability to anticipate user needs and act without explicit instruction.
For instance, based on explicit user request or implicit observations across multiple sessions, an agent might automatically run a specific type of linter after code changes or user a certain color palette for future user interface components.
While certainly useful for improving user experience and reducing friction, human-agent coding collaboration can function fully without proactivity: users can explicitly request every action, the agent responds, and the loop completes.
It enhances convenience but is not required for the fundamental interaction to succeed.

Effective proactivity emerges naturally from \textit{adaptability}.
While simple proactive behaviors can be implemented via static rules (e.g., always run tests after code changes), calibrated proactivity requires knowing when such actions are welcome versus intrusive.
An agent that learns a user's preferences, expertise level, and workflow patterns can determine when to take initiative versus wait for instruction~\citep{horvitz1999principles}.
Without such learned context, proactive actions risk being unwelcome interruptions rather than helpful assistance.

\textbf{Safety.}
Safety refers to preventing harmful, unintended, or irreversible agent behavior.
For instance, an agent should not delete critical files without confirmation, execute unvetted external code, or make sweeping changes that corrupt a codebase.
But this is also why we find that the definition of safety is difficult to formalize as a universal property, because its definition depends heavily on context.
With regard to the given example, it could also be argued that a file deletions may be desirable in one workflow (automated cleanup scripts), but catastrophic in another (production deployment).

We believe that safety failures can best be thought of as a product of breakdowns in existing pillars rather than the absence of a standalone safety mechanism.
Failures may arise from multiple interacting gaps: a file deletion gone wrong due to insufficient \textit{steerability} (no checkpoint before an irreversible action), while a corrupted codebase may reflect insufficient \textit{verification} (output not assessed before commit)~\citep{leveson2016engineering}.
When these pillars function well, agents that behave respectfully with respect to human expectations emerges as a natural, desirable side effect.

\textbf{Orchestration.}
Orchestration refers to the coordination of multiple agents and humans working together on shared or interdependent tasks.
For instance, a team of developers might delegate subtasks to different specialized agents, or multiple agents might collaborate on a large refactoring effort while a human architect provides high-level oversight.

Our position focuses on the fundamental dyad of a single human and a single coding agent.
While we agree that orchestration's potential and rich additional challenges (delegation, conflict resolution, group/pluralistic alignment) are worth investigating, it is likely that successful orchestration should be founded upon the pillars for our basic, dyadic setting functioning well.
Just as distributed systems build upon well-defined node-level protocols, we view orchestration as extending the interaction primitives discussed in this work to multi-party settings.
Effective multi-party collaboration requires that each dyad achieves task alignment, that handoffs preserve steerability, that distributed outputs remain verifiable, and that agents adapt to broader team conventions.

We leave an extra note that there are certainly several works for multi-agent coordination for coding and software engineering~\citep{hong2024metagptmetaprogrammingmultiagent,he2025llmbasedmultiagentsystemssoftware}.
However, prior works have overwhelmingly explored collaboration settings that don't involve human participation.

\textbf{Customizability.}
Customizability refers to a user's ability to manually configure agent behavior through skills, plugins, hooks, rules files, or other extensions.
For instance, users might add custom linting rules, define project-specific commands, or install plugins that integrate with their preferred toolchain.

Customizability is a helpful system feature that achieves outcomes similar to adaptability, but through explicit user effort rather than agent learning.
The core interaction loop functions without user-defined extensions; they enhance the experience but do not enable it.
Note that customizability differs from \textit{steerability} in several manners, most notably in temporal scope.
Steerability concerns in-the-moment control during task execution, while customizability involves adding persistent configuration that spans sessions.
In this sense, customizability and \textit{adaptability} are complementary: customizability provides immediate, user-controlled adjustments, while adaptability enables agents to internalize patterns over time without repeated manual configuration.



\textbf{Usability.}
Usability concerns how easily users can interact with an agent system, including interface clarity, interaction friction, and opportunities for learning.
For instance, whether a coding agent displays diffs inline or in a separate panel is a usability decision.
Although essential for real-world adoption, these factors are largely determined by design choices rather than underlying interaction capabilities, and we therefore view usability as complementary to the core pillars.

\subsection{On the Horizon}
\label{appx:on-the-horizon}

There is a potential disillusionment that readers may feel about this work.
Are our proposals stepping stones to a more collaboration-oriented future?
Or are we simply conceptualizing listening devices to elicit human signals for improving fundamentally autonomous models?
The second interpretation implies that our framework could accelerate knowledge worker displacement, reducing humans to transient reward models for systems designed to eventually replace them.

The truth is, we don't know.
Our argument is narrower: current research optimizes for full autonomy prematurely before we understand what effective collaboration looks like.
Whether full autonomy is the end state and how human-AI labor markets are affected en route are important but separate questions we aren't empirically equipped to answer.
That said, our opinion is that human involvement is not a stopgap, but intrinsically necessary.
For accountability, for judgment in novel situations, for alignment with intent only humans can ultimately adjudicate.

\end{document}